\documentclass[a4paper]{ESASPCS13Style}
\usepackage{epsfig}

\begin{document}

\title{Interpretation of the Spectra Originating from the Photospheres
Contaminated with Dust - Experience in L and T Dwarfs}
	
\author{T. Tsuji  \institute{Institute of Astronomy, School of 
Science, The University of Tokyo, Mitaka, Tokyo, 181-0015 Japan } }

\maketitle 

\begin{abstract}

  An essential feature of substellar dwarfs compared with the Sun
and stars is the formation of dust in their photospheres (not in the
outer envelope). It appears that
the observed data could be understood if the
dust exists in a form of thin cloud deep in the photosphere rather than in
the cooler surface region. Recent observations also show that the dust 
column densities in the observable photosphere are quite 
different for the same effective temperature, gravity, and metallicity, but 
the reason for such a sporadic variation is unknown. 
Moreover, the effect of dust 
cloud is difficult to discriminate from those by other basic parameters such as
the effective temperature which also has significant effect on the dust column 
density. For this reason, the spectra of dusty dwarfs
were in fact mis-interpreted by ourselves and will be reanalyzed in this
contribution. Also, 
even the  spectral classification is not free from such a difficulty, 
as is evidenced by an odd ``brightening'' of $M_{\rm bol}$ plotted against the 
L and T types.
   
\end{abstract}

\section{Introduction}

It is only a decade ago that a genuine brown dwarf was discovered 
(\cite{nak95}) and our 
experience in interpreting the spectra of brown dwarfs is
still meager compared with the longer experience in the Sun and stars.
A new feature in ultracool dwarfs such as brown dwarfs is the presence of 
dust in their photospheres, 
the possible importance of which has been recognized at an
early time (\cite{tsu96}). But how to treat the dust formed in the
photospheric environment was not known. If
a simple thermodynamical equilibrium is assumed, dust certainly forms
but how to sustain the dust in the photosphere was
unknown and  dust also tends to be over-produced compared to the known
observations in general. Actually, dust in the observable photosphere should 
be controlled by such processes as nucleation, growth, segregation, 
precipitation, evaporation etc., and we 
proposed a simple model referred to as the unified cloudy model (UCM) to take
into account these processes semi-empirically (\cite{tsu02}).

In the UCM, we assumed that dust forms at the condensation temperature
$ T_{\rm cond}$,  but dust soon grows to be too large at a slightly lower
temperature which we referred to as the critical temperature 
 $ T_{\rm cr}$ and segregates from the gaseous mixtures.
Thus dust forms a homogeneous cloud in the region where 
$  T_{\rm cr} \la T \la T_{\rm cond}$. In this model,
$ T_{\rm cond}$ is essentially determined by the thermodynamical data
but $ T_{\rm cr} $ is not predictable at present.
At first, we assumed that
$ T_{\rm cr} $ remains the same throughout L and T dwarfs for simplicity. 
Recent observations, however, revealed that $ T_{\rm cr}$, which
is a measure of the thickness of the cloud (a larger 
deviation of $ T_{\rm cr} $ from $ T_{\rm cond}$ implies a thicker cloud),
 shows a sporadic variation even at the fixed effective temperature (Sect.2).
Our UCM is already formulated to allow for the change of $ T_{\rm cr}$  and 
its application to the new situation is straightforward. 
Compared with our previous interpretation of the spectra of 
L and T dwarfs based on a constant value of $ T_{\rm cr}$,
our revised analysis allowing for the variation of $ T_{\rm cr}$ in UCMs  
results in a drastic change in our understanding of L and T dwarfs.

\section{ Infrared Colors}
\label{sec:ex}

It has been known that the infrared colors plotted against the spectral
types show the red limit at late L dwarfs (\cite{leg02}) and this may be
because the dust column density in the observable photosphere is the
largest in late L dwarfs. Although there were some scatters in the 
observed infrared colors, it appeared to be explained by our UCMs with
 $ T_{\rm cr} \approx 1800 \pm 100$\,K (\cite{tsu02}) and we assumed
a constant value of $ T_{\rm cr} = 1800$\,K in our further applications
of the UCMs (e.g. \cite{tsu04}). 

Recent observations, however, revealed that the infrared colors 
(\cite{kna04}) show a large variation if they are
plotted against  $T_{\rm eff}$ based on the bolometric fluxes 
(\cite{leg02}, \cite{gol04}) and parallaxes (\cite{vrb04}).
The case of $J-K$ is shown in Fig.1. The variation may not be explained
by the effect of log\,$g$ and metallicity. The predicted values of $J-K$ 
for $ T_{\rm cr} = 1700, 1800, 1900$\,K and $ T_{\rm cond}$ (this case
implies that dust disappears as soon as it is formed and thus there is
effectively no dust) are overlaid on Fig.1.   The variations of 
$J-K$ at a fixed $T_{\rm eff}$ are quite large, especially at around  
$T_{\rm eff} = 1400 \pm 100$\,K, and it is clear that the assumption of
a constant $T_{\rm cr} $ can no longer be supported.

\begin{figure}[ht]
\begin{center}  
\hspace{-5mm}  
\epsfig{file=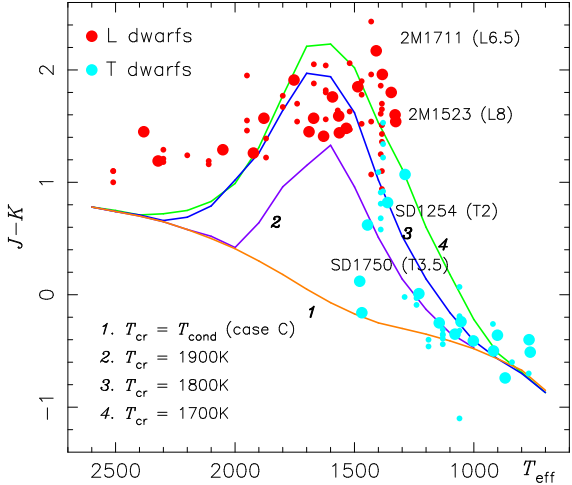,width=7.5cm}
\caption{Observed $J-K$   and predicted ones for different 
$T_{\rm cr}$ values plotted against $T_{\rm eff}$ 
(large and small circles distinguish the $T_{\rm eff}$ values by the direct 
determinations and by estimations based on the $T_{\rm eff}$ - Sp.\,Type 
relation, respectively). 
}
\label{Fig.1}
\end{center}
\end{figure}

\begin{figure}[hb]
\begin{center}  
\hspace{0mm}  
\epsfig{file=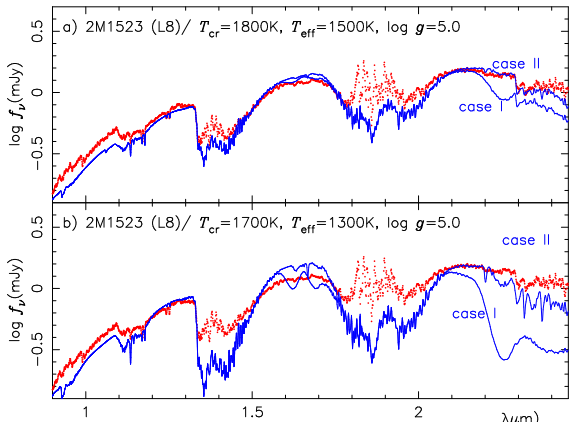,width=7.5cm}
\vspace{-2mm} 
\caption{Spectrum of 2MASS\,1523 (L8) and predictions based on UCMs of: 
a) $T_{\rm cr} = 1800$\,K
\&  $T_{\rm eff} = 1500$\,K (log\,$g$ = 5.0). b) $T_{\rm cr} = 1700$\,K
\&  $T_{\rm eff} = 1300$\,K (log\,$g$ = 5.0). 
}
\label{Fig.2}
\end{center}
\end{figure}

\section{Degeneracy of $T_{\rm eff}$ and $T_{\rm cr}$ on the Spectra}

 The dust column density generally increases at lower $T_{\rm eff}$
(at fixed $T_{\rm cr}$) and also at lower $T_{\rm cr}$ (at fixed 
$T_{\rm eff}$). For this reason, the effects of $T_{\rm eff}$ and $T_{\rm cr}$
on the spectra so to speak degenerate and are difficult to discriminate
unless one of them can be known by other methods. 
As an example, we reproduce the analysis of the L8 dwarf 2MASS\,1523 based on 
$T_{\rm cr} = 1800$\,K (\cite{tsu04}) in Fig.2a,  
showing that this L8 dwarf can be accounted for by 
the UCM with $T_{\rm eff} = 1500$\,K  (Cases I and II are based
on the CH$_4$ opacities with the band model method and linelist,
respectively).	  
However, recent infrared photometry  revealed that $T_{\rm cr} 
\approx 1700$\,K for 2MASS\,1523 (Fig.1) and we analyzed the same spectra 
based on this $T_{\rm cr}$ value.  The result
shown in Fig.2b reveals that 
the UCM with $T_{\rm cr} = 1700$\,K and $T_{\rm eff} = 1300$\,K provides a 
reasonable fit except for the water band regions. 
Thus different combinations of $T_{\rm eff}$ and $T_{\rm cr}$ could
explain the same spectrum so far as it is analyzed as a relative
spectral energy distribution (SED).  
 
\begin{figure}[hb]
\begin{center}  
\hspace{-5mm}  
\epsfig{file=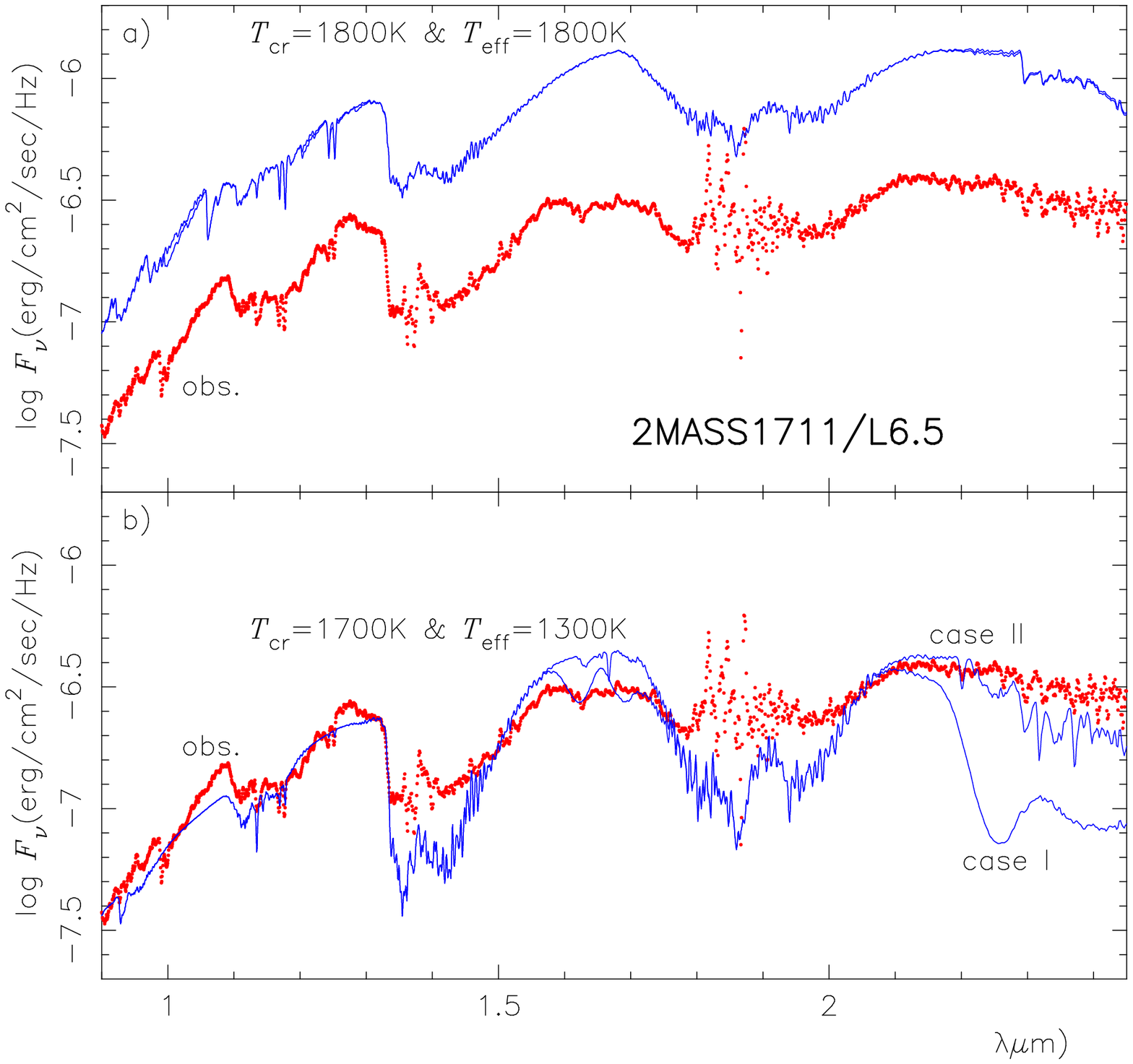,width=7.5cm}
\vspace{-2mm} 
\caption{Observed SED of 2MASS\,1711 (L6.5) reduced to an absolute scale 
(dots) is compared with the predicted ones (solid lines) characerized by: 
a) $T_{\rm cr} = 1800$\,K and $T_{\rm eff} = 1800$\,K (log\,$g$ = 5.0). 
b) $T_{\rm cr} = 1700$\,K and $T_{\rm eff} = 1300$\,K (log\,$g$ = 5.0).
}
\label{Fig.3}
\end{center}
\end{figure}

\begin{figure}[ht]
\begin{center}  
\hspace{-5mm}  
\epsfig{file=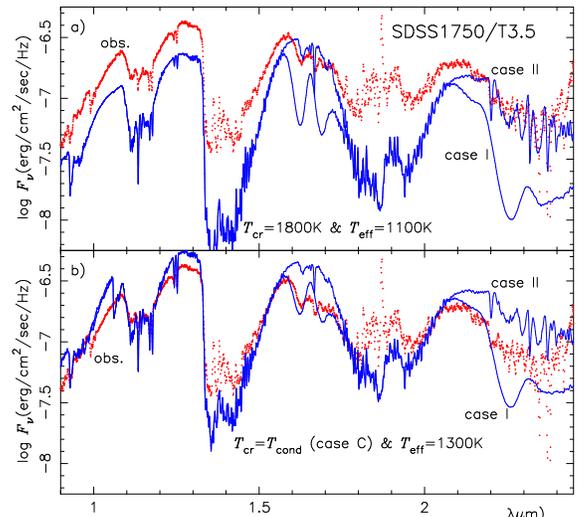,width=7.5cm}
\vspace{-2mm} 
\caption{Observed SED of SDSS\,1750 (T3.5) reduced to an absolute scale 
(dots) is compared with the predicted ones (solid lines) characterized by: 
a) $T_{\rm cr} = 1800$\,K and $T_{\rm eff} = 1100$\,K (log\,$g$ = 5.0). 
b) $T_{\rm cr} = T_{\rm cond}$      and $T_{\rm eff} = 1300$\,K (log\,$g$ = 
5.0). 
}
\label{Fig.4}
\end{center}
\end{figure}

\begin{table*} [ht]
\hspace{-40mm}
\caption{Effective temperatures  based on the UCMs with different $T_{\rm cr}$
values and empirical effective temperatures.  
}
\begin{center}
    \footnotesize
\begin{tabular}{ l l l l l l l}
\hline \hline
Object & Sp.type & $J-K$ (MKO) & $J-K$ (CIT) & $T_{\rm eff}~~ (T_{\rm cr})$ 
& $T_{\rm eff}~~(T_{\rm cr})$ &$T_{\rm eff}$~ (empirical)  \\
   &  &   Knapp et al.(2004) & Vrba et al.(2004)  & Tsuji et al.(2004) & 
Present results & Vrba et al.(2004)   \\  \hline 
2MASS\,1711 & L6.5 & - & 2.25  & 1800\,K (1800\,K) & 1300\,K (1700\,K) & 1545\,K \\   
2MASS\,1523 & L8 & 1.60 & 1.65  &  1500\,K (1800\,K) &  1300\,K (1700\,K) & 1330\,K \\   
SDSS\,1254 & T2  & 0.82 & 0.96 & 1300\,K (1800\,K) & 1300\,K (1800\,K) & 1361\,K \\   
SDSS\,1750 & T3.5 & 0.12 & 0.83 & 1100\,K (1800\,K) & 1300\,K ($T_{\rm cond}$) & 1478\,K  \\   
\hline \hline
\end{tabular}
\end{center}
\end{table*}

\section{How to Analyze the Spectra}

The ambiguity due to the degeneracy of $T_{\rm eff}$ and $T_{\rm cr}$
can be removed to some extend by transforming
the observed spectra to the SEDs on an absolute scale with the use
of the measured  parallaxes and assuming the Jupiter radius. 
As an example, it is immediately clear that the spectrum of 
2MASS\,1711 (L6.5) reduced to the emergent
flux on an absolute scale (in unit of erg\,cm$^{-2}$\,sec$^{-1}$\,Hz$^{-1}$)
 cannot be fitted with the predicted spectrum based 
on the UCM with  $T_{\rm cr} = 1800$\,K and $T_{\rm eff} = 1800$\,K (Fig.3a), 
even though the observed and predicted spectra can be fitted on the relative 
scale (i.e. by the shapes of the spectra) as done previously 
(\cite{tsu04}). The same observed spectrum is fitted better with  the predicted
one based on the UCM with  $T_{\rm cr} = 1700$\,K  and $T_{\rm eff} = 
1300$\,K (Fig.3b), both on the absolute and relative scales.
Note that $J-K$ (Fig.1) suggests even a lower value of $T_{\rm cr}$.
Thus the analysis of the SED on the absolute scale could
discriminate the different possible combinations of $T_{\rm eff}$ and 
$T_{\rm cr}$.

As another example, the SED of SDSS\,1750 (T3.5) on an 
absolute scale  can be
fitted only marginally with the prediction based on the UCM
with  $T_{\rm cr} = 1800$\,K and $T_{\rm eff} = 1100$\,K
(Fig.4a),   and the predicted water bands appear to be too strong.  
 The same observed SED  is compared with
the predicted one based on the UCM with  $T_{\rm cr} = T_{\rm cond}$ and
 $T_{\rm eff} = 1300$\,K (Fig.4b). 
The observed and predicted SEDs now
show better agreement both on the absolute and relative scales.
Note that the infrared color suggests 
$T_{\rm cr} \approx T_{\rm cond}$ for this object (Fig.1).

\section{How to Interprete the Spectral Sequence}
By the application of the UCMs  assuming a fixed value of
$T_{\rm cr} = 1800$\,K throughout, the spectra of cool dwarfs from 
L6.5 to T3.5 shown in Fig.5
could have been interpreted as a temperature sequence extending from 1800\,K
to 1100\,K, as is reproduced in the 5-th column of Table 1 (\cite{tsu04}). 
The result of our reanalysis based on the SEDs on the absolute scale
as outlined in Sect.4 is summarized in the 6-th column of Table 1.
The new result shows a drastic difference as compared with the previous one:
The spectral sequence extending form L6.5 to T3.5 is nothing to
do with $T_{\rm eff}$ but can be interpreted as the effect of 
$T_{\rm cr} $ alone.

Now a problem is which is a correct solution. We prefer the new
solution by the following reasons: First, the variable $T_{\rm cr} $ is
more consistent with the recent observations of the infrared colors (Fig.1).
Second, the resulting values of $T_{\rm eff} $ are more close to the recent
empirical values reproduced in the 7-th column of Table 1 (\cite{vrb04}).
Third, it is evident that the analysis of the SEDs on the absolute 
scale should be preferred if possible, and the analysis of the SEDs on
the relative scale (or the shape of the spectra) is misleading especially
because of the degeneracy between $T_{\rm eff} $ and $T_{\rm cr} $  as shown 
in Sect.3.

In conclusion, the L - T spectral sequence shown in Fig.5 is not  
a temperature sequence but is due to the change of $T_{\rm cr} $ and 
hence of  the thickness of the dust cloud.   This conclusion is quite surpring 
in that $T_{\rm eff}$ plays little role in such a 
distinct change of spectra requiring the different
spectral types L and T. This unexpected result is entirely due to dust,
which should be more important than has
been thought before. Also a parameter that specifies the nature
of the dust cloud, $T_{\rm cr}$ in our UCM, is sometimes more important
than $T_{\rm eff}$ in the characterization of cool dwarfs.
Thus we confirm that $T_{\rm cr}$ should be
regarded as a basic parameter together with $T_{\rm eff}$ and log\,$g$.  

\begin{figure}[hb]
\begin{center}  
\hspace{-5mm}  
\epsfig{file=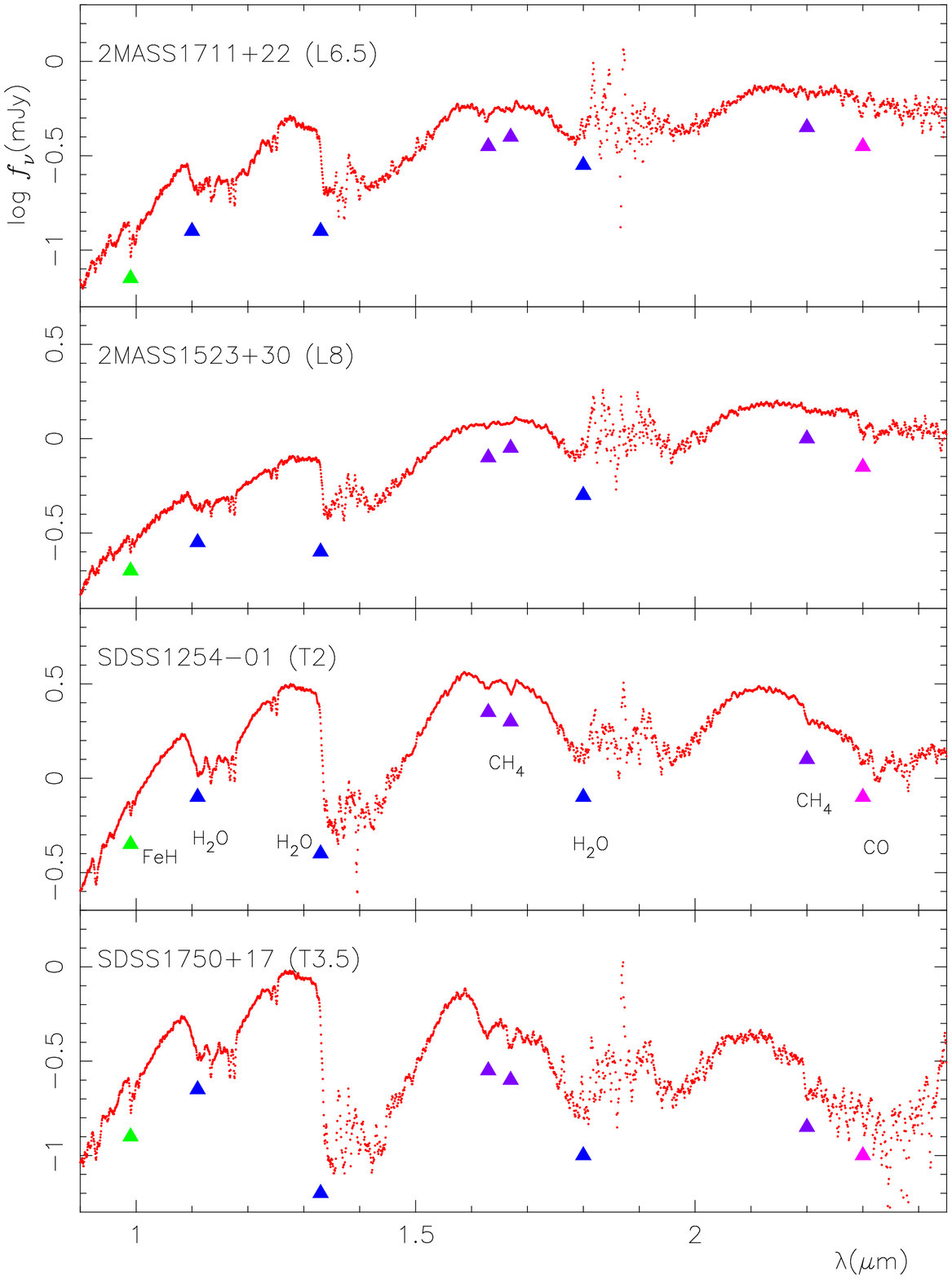,width=7.5cm}
\caption{Spectral sequence from L to T which, however, is not a sequence 
of $T_{\rm eff}$ but of $T_{\rm cr}$ or of dust column density (Sect.5).
}
\label{Fig.5}
\end{center}
\end{figure}

\begin{figure}[ht]
\begin{center}  
\hspace{-5mm}  
\epsfig{file=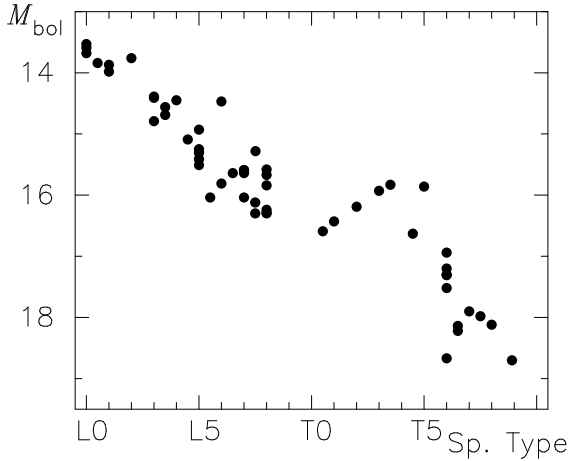,width=7.5cm}
\vspace{0mm} 
\caption{$M_{\rm bol}$ (Vrba et. al. 2004) plotted against L-T  types.
An odd brightening indicates that the L-T spectral sequence may not be 
a temperature sequence, at least partly.
}
\label{Fig.6}
\end{center}
\end{figure}

\section{The``$J$-Brightening'' in the Spectral Sequence}
It is known that the absolute $J$ magnitude plotted against the L-T
type shows an odd brightening at early T types (\cite{dah02},
\cite{tin03}, \cite{vrb04}). It was suspected that this phenomenon may be
due to some atmospheric effect, but no model including our UCM could explain 
this observation at all. The brightening is also observed in the $H$ and $K$ 
bands if not so pronounced as in the $J$ band. But then the absolute bolometric
magnitude, which largely depends on the $J, H,$ and $K$ fluxes, should also 
show the brightening and this is in fact found to be the case as shown
in Fig.6. Since the L and T dwarfs are evolving on the cooling tracks, their 
bolometric luminosities should never show ``brightening'' if they are plotted 
againt a correct temperature indicator.
This result implies that the L-T spectral sequence may not be a temperature 
sequence, consistent with the conclusion of Sect.5, and
the ``$J$-brightening'' as well as the  ``$M_{\rm bol}$-brightening'' may  
simply be an artifact of the L-T spectral classification, in which the
effects of $T_{\rm eff}$ and $T_{\rm cr}$ are mixed as shown in Sect.3. 
Thus the so-called ``$J$-brightening'' problem is solved or, more properly,
this problem now disappeared. Instead a more serious problem of how to 
classify the spectra of ultracool dwarfs stands before us.

\section{Discussion and Concluding Remarks}
\label{sec:tit}

In the interpretation of the spectra of dust-contaminated object
such as brown dwarfs, a parameter that specifies the thickness of the
dust cloud, $T_{\rm cr}$ in our formulation or $f_{\rm sed}$ 
in that of \cite*{mar02}, plays an important role.
At present, it seems to be difficult to predict the value of $T_{\rm cr}$
based on basic physics, especially because it is sporadic rather than
related to the other basic parameters (Sect.2).    
The variation of $T_{\rm cr}$ is especially large at $T_{\rm eff} \approx
1400$\,K where the second convective zone appears in addition to the
one deep in the photosphere (\cite{tsu02}). It may be possible that
$T_{\rm cr}$ is related to the convective activities, but details are yet
to be explored.
Thus, in addition to the four parameters generally
required to specify the stellar photosphere, i.e.,  chemical composition, 
 $T_{\rm eff}$, log\,$g$,  and micro-turbulent velocity, fifth parameter
$T_{\rm cr}$ is needed for the characterization  of dusty dwarfs. 

It is to be noted that the 
turbulent velocity is still determined empirically since its 
discovery more than half a century ago (\cite{str34}) and not yet 
predictable based on basic physics for individual objects.
At present, we must leave $T_{\rm cr}$ as an empirical parameter something
like the turbulent velocity, but
there are difficulties inherent to dust. For example, dust, 
unlike atoms and molecules, shows almost no clear spectral signature and it 
is difficult to estimate the dust column density directly. While dust plays 
significant role in defining the spectral characteristics, spectral 
classification had to be done on the spectral features originating
from atoms and molecules (e.g. \cite{kir99}, \cite{mar99}, \cite{bur02}, 
\cite{geb02}), and such a difficulty should be fatal in the spectral 
classification of dusty objects. 

Despite some formidable problems, recent progress in observation of such
faint objects as brown dwarfs, not only in spectroscopy but also in 
photometry and astrometry,
is quite marvelous, and a more consistent interpretation can be 
achieved by considering all these data collectively.  




\begin{thebibliography}{}
\bibitem[\protect
\astroncite{Burgasse et~al.}{2002}]{bur02}
Burgasser, A. J., Kirkpatrick, J. D., Brown, M. E. et al.  2002, ApJ, 564,
421 

\bibitem[\protect
\astroncite{Dahn  et~al.}{2002}]{dah02} 
Dahn, C. C., Harris, H. C., Vrba, F. J. et al.\ 2002, AJ, 124, 1170 

\bibitem[\protect
\astroncite{Geballe  et~al.}{2002}]{geb02} 
Geballe, T. R., Knapp, G. R., Leggett, S. K. et al.\ 2002, ApJ, 564, 466

\bibitem[\protect
\astroncite{Golimowski et~al.}{2004}]{gol04}
Golimowski, D. A., Leggett, S. K., Marley, M. S. et al.\ 2004, AJ, 127, 3516 

\bibitem[\protect
\astroncite{Kirkpatrick et~al.}{1999}]{kir99}
Kirkpatrick, J. D., Reid, I. N., Liebert, J. et. al. 1999, ApJ, 519, 802

\bibitem[\protect
\astroncite{Knapp et~al.}{2004}]{kna04}
Knapp., G. R., Leggett, S. K., Fan, X. et al. 2004, AJ, 127, 3553 

\bibitem[\protect
\astroncite{Leggett et~al.}{2002}]{leg02} 
Leggett, S. K., Golimowski, D. A., Fan, X. et al. 2002, ApJ, 564, 452 

\bibitem[\protect
\astroncite{Marley et~al.}{2002}]{mar02}
Marley, M. S., Seager, S., Saumon, D. et al. 2002, ApJ, 568, 335

\bibitem[\protect
\astroncite{Mart\'in et~al.}{1999}]{mar99}
Mart\'in, E. L., Delfosse, X., Basri, G. et al. 1999, AJ, 118, 2466

\bibitem[\protect
\astroncite{Nakajima et~al.}{1995}]{nak95}
Nakajima, T., Oppenheimer, B. R., Kulkarni, S. R. et al. 1995, Nature, 378, 463

\bibitem[\protect
\astroncite{Struve \& Elvey}{1934}]{str34}
Struve, O., \& Elvey, C. T. 1934, ApJ, 79, 409

\bibitem[\protect
\astroncite{Tinney et~al.}{2003}]{tin03}
Tinney, C. G., Burgasser, A. J., Kirkpatrick, J. D. 2003, AJ, 126, 975 

\bibitem[\protect
\astroncite{Tsuji}{2002}]{tsu02}
Tsuji, T. 2002, ApJ, 575, 264  

\bibitem[\protect
\astroncite{Tsuji et~al.}{2004}]{tsu04}
Tsuji, T., Nakajima, T., Yanagisawa, K. 2004, ApJ, 607, 511  

\bibitem[\protect
\astroncite{Tsuji et~al.}{1996}]{tsu96}
Tsuji, T., Ohnaka, K., Aoki, W. 1996a, A\&A, 305, L1 

\bibitem[\protect
\astroncite{Vrba et~al.}{2004}]{vrb04}
Vrba, F. J., Henden, A. A., Luginbuhl, H. H. et al. 2004, AJ, 127, 2948  

\end{thebibliography}
\end{document}